\documentclass{ws-procs9x6}

\usepackage{epsfig,graphics}

\newcommand{\tr}{\rm tr \,}

\newcommand{\TableHeader}[0]{\begin{tabular}{|c|c|c|c|c|c|} \hline (I,S)  &  Kanal &  Lage &  |g|  & Lage  &|g| \\}
\def\rescale{\fontsize{6}{1}}

\begin{document}

\title{Dynamically generated baryon resonances}

\author{M.F.M. Lutz and J. Hofmann}

\address{Gesellschaft f\"ur Schwerionenforschung (GSI)\\
Planck Str. 1, 64291 Darmstadt, Germany}

\maketitle

\abstracts{Identifying a zero-range exchange of vector mesons as
the driving force for the s-wave scattering of pseudo-scalar
mesons off the  baryon ground states, a rich spectrum of molecules
is formed. We argue that chiral symmetry and large-$N_c$
considerations determine that part of the interaction which
generates the spectrum. We suggest the existence of strongly bound
crypto-exotic baryons, which contain a charm-anti-charm pair. Such
states are narrow since they can decay only via OZI-violating
processes. A narrow nucleon resonance is found at mass 3.52 GeV.
It is a coupled-channel bound state of the $(\eta_c\,N), (\bar
D\,\Sigma_c)$ system, which decays dominantly into the $(\eta' N)$
channel. Furthermore two isospin singlet hyperon states at mass
3.23 GeV and 3.58 GeV are observed as a consequence of
coupled-channel interactions of the $(\bar D_s\,\Lambda_c), (\bar
D\,\Xi_c)$ and $(\eta_c \,\Lambda),(\bar D\,\Xi_c')$ states. Most
striking is the small width of about 1 MeV of the lower state. The
upper state may be significantly broader due to a strong coupling
to the $(\eta' \Lambda)$ state. The spectrum of crypto-exotic
charm-zero states is completed with an isospin triplet state at
3.93 GeV and an isospin doublet state at 3.80 GeV. The dominant
decay modes involve again the $\eta'$ meson.}

\section{Introduction}

\begin{figure}[t]
\epsfxsize=13.5cm   
\epsfbox{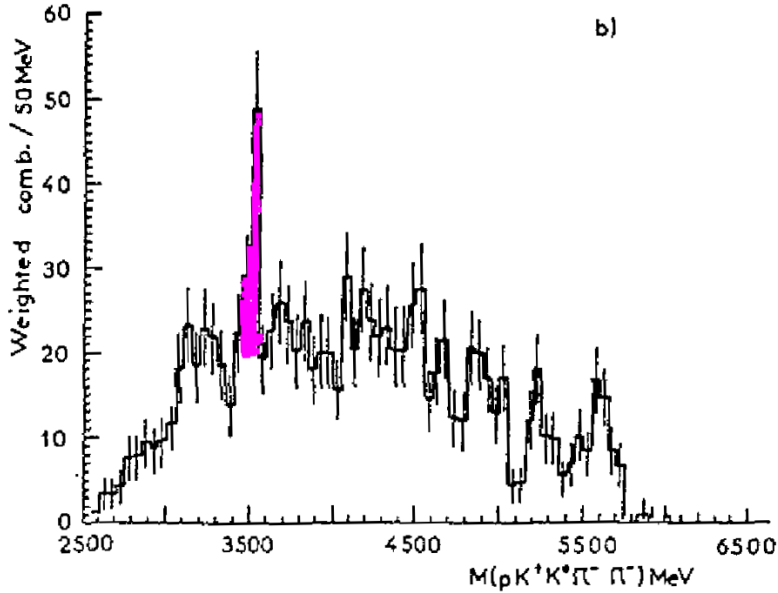}
\caption{Measured five-body invariant mass distribution in a 19
GeV $\pi^-$ beam experiment at CERN. The figure is taken from Ref.$\,^6$}
\label{fig:Nstar}
\end{figure}

The existence of strongly bound crypto-exotic baryon systems with
hidden charm would be a striking feature of strong interactions
\cite{Brodsky:Schmidt:Teramond:90,Kaidalov:Volkovitsky:92,Gobbi:Riska:Scoccola:92}.
Such states may be narrow since their strong decays are
OZI-suppressed \cite{Landsberg:94}. There are experimental hints
that such states may indeed be part of nature. A high statistics
bubble chamber experiment performed 30 years ago with a $K^-$ beam
reported on a possible signal for a hyperon resonance of mass 3.17
GeV of width smaller than 20 MeV \cite{Amirzadeh:79}.  About ten
years later a further bubble chamber experiment using a high
energy $\pi^-$ beam suggested a nucleon resonance of mass 3.52 GeV
with a narrow width of $7^{+20}_{-7}$ MeV . In Fig.
\ref{fig:Nstar} we recall the measured five body
($p\,K^+\,K^0\,\pi^-\pi^-$) invariant mass
distribution\cite{Karnaukhov:91} suggesting the existence of a
crypto exotic nucleon resonance.

It is the purpose of the present talk to review a study addressing
the possible existence of crypto-exotic baryon
systems\cite{Hofmann:Lutz:2005}. In view of the highly speculative
nature of such states it is important to correlate the properties
of such states to those firmly established, applying a unified and
quantitative framework. We extended previous works
\cite{LK04-charm,LK05} that performed a coupled-channel study of
the s-wave scattering processes where a Goldstone boson hits an
open-charm baryon ground state. The spectrum of
$J^P=\frac{1}{2}^-$ and $J^P=\frac{3}{2}^-$ molecules obtained in
\cite{LK04-charm,LK05} is quite compatible with the so far very
few observed states. Analogous computations successfully describe
the spectrum of open-charm mesons with $J^P=0^+$ and $1^+$ quantum
numbers \cite{KL04,HL04}. These developments were driven by the
hadrogenesis conjecture: meson and baryon resonances that do not
belong to the large-$N_c$ ground state of QCD should be viewed as
hadronic molecular states
\cite{LK02,LWF02,LK04-axial,Granada,Copenhagen}. Generalizing
those computations to include D- and $\eta_c$-mesons in the
intermediate states offers the possibility to address the
formation of crypto-exotic baryon states (see also
\cite{Mo:80,Rho:Riska:Scoccola:92,Min:Oh:Rho:95,Oh:Kim:04}).

The results of \cite{LK04-charm,LK05} were based on the leading
order chiral Lagrangian, that predicts unambiguously the s-wave
interaction strength of Goldstone bosons with open-charm baryon
states in terms of the pion decay constant. Including the light
vector mesons as explicit degrees of freedom in a chiral
Lagrangian gives an interpretation of the leading order
interaction in terms of the zero-range t-channel exchange of light
vector mesons \cite{Weinberg:68,Wyld,Dalitz,sw88,Bando:85}. The
latter couple universally to any matter field in this type of
approach. Based on the assumption that the interaction strength of
D- and $\eta_c$-mesons with the baryon ground states is also
dominated by the t-channel exchange of the light vector mesons, we
performed a coupled-channel study of crypto-exotic baryon
resonances\cite{Hofmann:Lutz:2005}.

\section{Coupled-channel interactions}

We consider the interaction of pseudoscalar mesons with the
ground-state baryons composed out of u,d,s,c quarks. The
pseudoscalar mesons that are considered in this work can be
grouped into multiplet fields $\Phi_{[9]}, \Phi_{[\bar 3]}$ and
$\Phi_{[1]}$ corresponding to the Goldstone bosons, the $\eta'$,
the $D$-mesons and the $\eta_c$. The baryon states are collected
into SU(3) multiplet fields $B_{[8]},B_{[6]}$ and $B_{[\bar 3]}$
with charm 0,1 and 1. For an explicit representation of the
various multiplet fields we refer to \cite{Hofmann:Lutz:2005}.

In a first step we construct the interaction of the mesons and
baryon fields with the nonet-field $V_{[9]}^\mu$ of light vector
mesons. We write down a list of relevant SU(3) invariant 3-point
vertices that involve the minimal number of derivatives. Consider
first the terms involving pseudo-scalar fields:
\begin{eqnarray} \label{LM}
\mathcal{L}^{\rm{SU(3)}}_{\rm{int}}
&=& {\textstyle{i\over 4}}\, h_{\bar 3 \bar 3}^9\    {\rm tr}\left(
 (\partial_{\mu}\Phi_{[\bar 3]})\, \Phi^\dagger_{[\bar 3]} V_{[9]}^{\mu} -
  \Phi_{[\bar 3]}\,
(\partial_{\mu}\Phi^\dagger_{[\bar 3]})\,
V_{[9]}^{\mu}\right)
\nonumber\\
&+& {\textstyle{i\over 4}}\, h_{\bar 3 \bar 3}^1\    {\rm tr}\left(
(\partial_{\mu}\Phi_{[\bar 3]})\, \Phi^\dagger_{[\bar 3]}-
\Phi_{[\bar 3]}\,(\partial_{\mu}\Phi^\dagger_{[\bar 3]})  \right) \cdot {\rm tr} \left (V_{[9]}^{\mu} \right)
\nonumber\\
&+& {\textstyle{i\over 4}}\, h_{99}^9\    {\rm tr}\left(
(\partial_{\mu}\Phi_{[9]})\, \Phi_{[9]}\, V_{[9]}^{\mu}
-\Phi_{[9]}\,(\partial_{\mu} \Phi_{[9]})\,  V_{[9]}^{\mu} \right)\,.
\label{vector-pseudoscalar}
\end{eqnarray}
It needs to be emphasized that the terms in (\ref{vector-pseudoscalar}) are
not at odds with the constraints set by chiral symmetry provided the light vector
mesons are coupled to matter fields via a gauge principle \cite{Weinberg:68,Bando:85}.
The latter requires a correlation of the coupling constants $h$ in (\ref{vector-pseudoscalar})
\begin{eqnarray}
h_{99}^9 = \frac{(m^{(V)}_{[9]})^2}{2\,g\,f^2}\,, \qquad h_{\bar 3 \bar 3}^{9} = 2\,g \,,
\label{chiral-constraint-mesons}
\end{eqnarray}
with the pion decay constant $f \simeq 92$ MeV. Here the universal vector coupling strength is
$g \simeq 6.6$ and the mass of the light vector mesons is $m_{[9]}^{(V)}$.

We continue with the construction of the three-point vertices
involving baryon fields. A list of SU(3) invariant terms reads:
\begin{eqnarray}
\mathcal{L}^{\rm{SU(3)}}_{\rm{int}} &=& {\textstyle{1\over
2}}\,g_{\bar 3 \bar 3}^{9}\, {\tr }\Big(\bar B_{[\bar
3]}\,\gamma_\mu \,V_{[9]}^\mu\, B_{[\bar 3]} \Big)
+{\textstyle{1\over 2}}\,g_{\bar 3 \bar 3}^{1}\, {\tr }\Big(\bar
B_{[\bar 3]}\,\gamma_\mu \, B_{[\bar 3]} \Big)\, {\tr} \Big(
\,V_{[9]}^\mu \Big)
\nonumber\\
&+&{\textstyle{1\over 2}}\,g_{66}^{9}\,
{\tr }\Big(\bar B_{[6]}\,\gamma_\mu \,V_{[9]}^\mu\, B_{[6]} \Big)
+{\textstyle{1\over 2}}\,g_{66}^{1}\,
{\tr }\Big(\bar B_{[6]}\,\gamma_\mu \, B_{[6]} \Big)\, {\tr} \Big( \,V_{[9]}^\mu \Big)
\nonumber\\
&+&{\textstyle{1\over 2}}\,g_{88}^{9-}\,
{\tr }\Big(\bar B_{[8]}\,\gamma_\mu \,\Big[ V_{[9]}^\mu\,, B_{[8]}\Big]_- \Big)
+{\textstyle{1\over 2}}\,g_{88}^{9+}\,
{\tr }\Big(\bar B_{[8]}\,\gamma_\mu \,\Big[ V_{[9]}^\mu\,, B_{[8]}\Big]_+ \Big)
\nonumber\\
&+&{\textstyle{1\over 2}}\,g_{88}^{1}\,
{\tr }\Big(\bar B_{[8]}\,\gamma_\mu \, B_{[8]} \Big)\, {\tr} \Big( \,V_{[9]}^\mu \Big)
\nonumber\\
&+&{\textstyle{1\over 2}}\,g_{\bar36}^{9}\,
{\tr }\Big(\bar B_{[\bar3]}\,\gamma_\mu\, V_{[9]}^\mu \, B_{[6]}+ \bar B_{[6]}\,\gamma_\mu\, V_{[9]}^\mu \, B_{[\bar 3]} \Big)\,.
\label{vector-baryons}
\end{eqnarray}
Within the hidden local symmetry model \cite{Bando:85} chiral
symmetry is recovered with
\begin{eqnarray}
&& g_{\bar 3 \bar 3}^9 =g_{66}^{9}= 2\,g\,,
\qquad g_{88}^{9,-}= g\,, \qquad \! g_{88}^{9,+}= 0\,, \qquad g_{\bar36}^{9}=0 \,.
\label{chiral-constraint-baryons}
\end{eqnarray}
It is acknowledged that chiral symmetry does not constrain the coupling
constants in (\ref{vector-pseudoscalar}, \ref{vector-baryons}) involving the SU(3) singlet
part of the fields. The latter can, however, be
constrained  by a large-$N_c$ operator analysis \cite{DJM}.
At leading order in the $1/N_c$ expansion the OZI rule \cite{OZI} is predicted.  As
a consequence the estimates
\begin{eqnarray}
&& h^1_{\bar 3\bar 3} = -g\,,\qquad
 g^1_{\bar 3 \bar 3}=g^1_{66}=0 \,,\qquad  g^1_{88}= g\,,\quad
\label{OZI-constraint}
\end{eqnarray}
follow. We emphasize that the combination of chiral and large-$N_c$ constraints
(\ref{chiral-constraint-mesons}, \ref{chiral-constraint-baryons}, \ref{OZI-constraint})
determine all coupling constants introduced in (\ref{vector-pseudoscalar}, \ref{vector-baryons}).

We close this section by investigating the coupling of heavy vector mesons, $V^\mu_{[\bar 3]}$ and
$V_{[1]}^\mu$ to the meson and baryon fields.
First we construct the most general SU(3) symmetric interaction terms involving the
pseudoscalar fields:
\begin{eqnarray}
{\mathcal L}_{\rm{int}}^{\rm{SU(3)}} &=&
{\textstyle{i\over 4}}\, h_{\bar 3 \bar 3}^0\
{\rm tr}\left(
(\partial_{\mu}\Phi_{[\bar 3]})\, \Phi^\dagger_{[\bar 3]} V_{[1]}^{\mu}
-\Phi_{[\bar 3]}\,(\partial_{\mu}\Phi^\dagger_{[\bar 3]})\, V_{[1]}^{\mu}  \right)
\nonumber\\
&+& {\textstyle{i\over 4}}\, h_{9\bar 3}^{\bar 3}\
{\rm tr}\left(
(\partial_{\mu} \Phi^\dagger_{[\bar 3]})\, \Phi_{[9]}\, V_{[\bar 3]}^{\mu}-
\Phi_{[9]}\,(\partial_{\mu}\Phi_{[\bar 3]})\, V_{[\bar 3]}^{\dagger \mu}  \right)
\nonumber\\
&+& {\textstyle{i\over 4}}\, h_{\bar 39}^{\bar 3}\
{\rm tr}\left(
(\partial_{\mu}\Phi_{[9]})\, \Phi_{[\bar 3]} V_{[\bar 3]}^{\dagger \mu}-
\Phi^\dagger_{[\bar 3]}\,(\partial_{\mu}\Phi_{[9]})\, V_{[\bar 3]}^{\mu}  \right)
\nonumber\\
&+& {\textstyle{i\over 4}}\, h_{\bar 3 0}^{\bar 3}\
{\rm tr}\left(
(\partial_{\mu}\Phi_{[1]})\, \Phi_{[\bar 3]} V_{[\bar 3]}^{\dagger \mu} -
 \Phi^\dagger_{[\bar 3]}\,(\partial_{\mu}\Phi_{[1]})\, V_{[\bar 3]}^{ \mu} \right)
\nonumber\\
&+& {\textstyle{i\over 4}}\, h_{0\bar 3}^{\bar 3}\
{\rm tr}\left(
(\partial_{\mu}\Phi^\dagger_{[\bar 3]}) \,\Phi_{[1]} V_{[\bar 3]}^{\mu}
-\Phi_{[1]}\,(\partial_{\mu}\Phi_{[\bar 3]})\, V_{[\bar 3]}^{\dagger \mu}  \right)
\nonumber\\
&+& {\textstyle{i\over 4}}\, h_{\bar 3 1}^{\bar 3}\
{\rm tr}\left( \Phi^\dagger _{[\bar 3]}\,V_{[\bar 3]}^{\mu}-
\Phi_{[\bar 3]}\,V_{[\bar 3]}^{\dagger \mu}
\right)
{\tr }(\partial_{\mu}\Phi_{[9 ]})\,
\nonumber\\
&+& {\textstyle{i\over 4}}\, h_{1 \bar 3 }^{\bar 3}\
{\rm tr}\left( (\partial_{\mu} \Phi_{[\bar 3]})\,V_{[\bar 3]}^{\dagger\mu}-
(\partial_{\mu} \Phi^\dagger_{[\bar 3]})\,V_{[\bar 3]}^{ \mu}
\right)
{\tr }(\Phi_{[9 ]})\,,
\label{vector-heavy-meson}
\end{eqnarray}
where the SU(4) symmetric gerneralization of (\ref{LM}) suggests the identification
\begin{eqnarray}
&&h^0_{\bar 3 \bar 3} = \sqrt{2}\,g\,, \quad
h^{\bar 3}_{9 \bar 3} = 2\,g\,, \quad \;\;\,\,
h^{\bar 3}_{\bar 3 9} = 2\,g\,,
\nonumber\\
&&h^{\bar 3}_{\bar 3 0} = \sqrt{2}\,g\,, \quad
h^{\bar 3}_{0 \bar 3} = \sqrt{2}\,g\,, \quad
h^{\bar 3}_{\bar 3 1} = g\,, \quad \;
h^{\bar 3}_{1 \bar 3} = g \,.
\label{meson-SU4-result}
\end{eqnarray}
The prediction of the SU(4)-symmetric coupling constants 
can be tested against the decay pattern of the D-meson. From the
empirical branching ratio \cite{PDG04}
 we deduce $(h^{\bar 3}_{\bar 3 9}+h^{\bar 3}_{9\bar 3 })/4 = 10.4 \pm 1.4$,
which is confronted with the SU(4) estimate $(h^{\bar 3}_{\bar 3
9}+h^{\bar 3}_{9\bar 3 })/4=g \simeq 6.6 $. We observe a moderate
SU(4) breaking pattern. Based on this result one may expect the
relations (\ref{meson-SU4-result}) to provide magnitudes for the
coupling constants reliable within a factor two.

We close this section with the construction of the most general SU(3) symmetric interaction
involving baryon fields
\begin{eqnarray}
{\mathcal L}_{\rm{int}}^{\rm{SU(3)}} &=&
{\textstyle{1\over 2}}\ g_{88}^0 \ {\rm tr} \left( \bar B_{[8]}  \,\gamma_{\mu}\,  B_{[8]}  \,V_{[1]}^{\mu}  \right)
+ {\textstyle{1\over 2}}\ g_{66}^0  \ {\rm tr}\left( \bar B_{[6]}  \,\gamma_{\mu}\,  B_{[6]}  \,V_{[1]}^{\mu}  \right)
\nonumber\\
&+&{\textstyle{1\over 2}}\ g_{\bar 3\bar 3}^0 \ {\rm tr}\left( \bar B_{[\bar 3]}  \,\gamma_{\mu}\,  B_{[\bar 3]}  \,V_{[1]}^{\mu}  \right)
\nonumber\\
&+&{\textstyle{1\over 2}}\ g_{86}^{\bar 3}\   {\rm tr}\left( \bar B_{[8]} \,\gamma_{\mu}\, B_{[6]} \,
V_{[\bar 3]}^{\dagger \mu}
+  \bar B_{[6]} \,\gamma_{\mu}\, B_{[8]} \,V_{[\bar 3]}^{\mu} \right)
\nonumber\\
&+& {\textstyle{1\over 2}}\ g_{8\bar{3}}^{\bar 3}\ {\rm tr}\left( \bar B_{[8]} \,\gamma_{\mu}\, B_{[\bar 3]}
\,V_{[\bar 3]}^{\dagger \mu}  +
 \bar B_{[\bar 3]} \,\gamma_{\mu}\, B_{[8]} \,V_{[\bar 3]}^{\mu} \right) \,,
\label{vector-heavy-baryon}
\end{eqnarray}
where a SU(4) symmetric vertex implies
\begin{eqnarray}
&& g^0_{88 } = 0\,, \quad
g^{0}_{66} = \sqrt{2}\,g\,, \quad
g^{0}_{\bar 3 \bar 3} =\sqrt{2}\, g\,, \quad
 g^{\bar 3}_{86} =\sqrt{2}\, g\,, \quad
g^{\bar 3}_{8 \bar 3} =-\sqrt{6}\, g\,, \quad
\label{baryon-SU4-result}
\end{eqnarray}
Unfortunately there appears to be no way at present to check on
the usefulness of the result (\ref{baryon-SU4-result}). Eventually
simulations of QCD on a lattice may shed some light on this issue.
The precise values of the coupling constants
(\ref{meson-SU4-result}, \ref{baryon-SU4-result}) do not affect
the major results of this study. This holds as long as those
coupling constants range in the region suggested by
(\ref{meson-SU4-result}, \ref{baryon-SU4-result}) within a factor
two to three.

\section{S-wave baryon resonances with zero charm}

The spectrum of $J^P=\frac{1}{2}^- $ baryon resonances as
generated by the t-channel vector-meson exchange interaction via
coupled-channel dynamics falls into two types of states.
Resonances with masses above 3 GeV couple strongly to mesons with
non-zero charm content. In the SU(3) limit those states form an
octet and a singlet. All other states have masses below 2 GeV. In
the SU(3) limit they group into two degenerate octets and one
singlet. The presence of the heavy channels does not affect that
part of the spectrum at all. This is reflected in coupling
constants of those states to the heavy channels within the typical
range of $g \sim 0.1$ (see Tabs.
\ref{tab:charm0a}-\ref{tab:charm0b}). We reproduce the success of
previous coupled-channel computations \cite{Granada,Copenhagen},
which predicts the existence of the s-wave resonances $N(1535),
\Lambda(1405), \Lambda(1670),\Xi(1690)$ unambiguously with masses
and branching ratios quite compatible with empirical information.
There are some quantitative differences. This is the consequence
of the t-channel vector meson exchange, which, only in the SU(3)
limit with degenerate vector meson masses, is equivalent to the
Weinberg-Tomozawa interaction the computation in
\cite{Granada,Copenhagen} was based on.

Most spectacular are the resonances with hidden charm above 3 GeV.
The multiplet structure of such states is readily understood. The
mesons with $C=-1$  form a triplet which is scattered off the
$C=+1$ baryons forming an anti-triplet or sextet. We decompose the
products into irreducible tensors
\begin{eqnarray}
3 \otimes \overline 3 = 1 \oplus 8 \,, \qquad
3 \otimes 6 = 8\oplus 10\,.
\label{3times6}
\end{eqnarray}
The coupled channel interaction is attractive in the singlet for
the triplet of baryons. Attraction in the octet sector is provided
by the sextet of baryons. The resulting octet of states mixes with
the $\eta'\,(N,\Lambda, \Sigma, \Xi)$ and $\eta_c\,(N,\Lambda,
\Sigma, \Xi)$ systems. A complicated mixing pattern arises. All
together the binding energies of the crypto-exotic states are
large. This is in part due to the large masses of the
coupled-channel states: the kinetic energy the attractive
t-channel force has to overcome is reduced.

\begin{table}[t]
\begin{center}
\tbl{
Spectrum of $J^P=\frac{1}{2}^-$ baryons with charm zero. The 3rd and 4th columns follow
with SU(4) symmetric 3-point vertices. In the
5th and 6th columns SU(4) breaking is introduced with  $h_{\bar 3 \bar 3}^1  \simeq -1.19 \,g$ and
$h^{\bar 3}_{ \bar 3 1 }=h^{\bar 3}_{1 \bar 3  } \simeq 0.71 \,g $. We use $g=6.6$.}
{\rescale \setlength{\tabcolsep}{1.0mm}
\setlength{\arraycolsep}{2.mm}
\renewcommand{\arraystretch}{0.75}
\begin{tabular}{|ll|c|c|c|c|c|}
\hline $C=0:$ &$ (\,I,\phantom{+}S)$  &
$\rm state$ &  $\begin{array}{c} M_R [\rm MeV]  \\ \Gamma_R \,[\rm MeV]  \end{array}$ &
$|g_R|$  & $\begin{array}{c} M_R [\rm MeV]  \\ \Gamma_R \,[\rm MeV]  \end{array}$ & $|g_R|$  \\
\hline
\hline
&$(\frac12,\phantom{+}0)$   &
$\begin{array}{l}  \pi\, N \\ \eta \,N \\ K \,\Lambda \\ K \,\Sigma \\
\eta' N \\ \eta_c N \\ \bar{D} \,\Lambda_c \\ \bar{D}\, \Sigma_c \end{array}$
& $\begin{array}{c} 1535 \\ 95 \end{array}$ & $\begin{array}{c} 0.3 \\ 2.1 \\ 1.7 \\ 3.3 \\ 0.0 \\ 0.0 \\ 0.2 \\ 0.2 \end{array}$
& $\begin{array}{c} 1536 \\ 94 \end{array}$ & $\begin{array}{c} 0.3 \\ 2.1 \\ 1.6 \\ 3.3 \\ 0.0 \\ 0.0 \\ 0.2 \\ 0.2 \end{array}$\\
\cline{3-7}
&$(\frac12,\phantom{+}0)$   &
$\begin{array}{l}  \pi\, N \\ \eta\, N \\ K \,\Lambda \\ K \,\Sigma \\
\eta' N \\ \eta_c N \\ \bar{D}\, \Lambda_c \\ \bar{D}\, \Sigma_c \end{array}$
& $\begin{array}{c} 3327 \\ 156 \end{array}$ & $\begin{array}{c} 0.1  \\ 0.1  \\ 0.1  \\ 0.1  \\ 1.4  \\ 0.7 \\ 0.5  \\ 5.7 \end{array}$
& $\begin{array}{c} 3520 \\ 7.3 \end{array}$ & $\begin{array}{c} 0.07 \\ 0.11 \\ 0.08 \\ 0.08 \\ 0.22 \\ 1.0 \\ 0.05 \\ 5.3 \end{array}$\\
\hline
&$(0,-1)$    &
$\begin{array}{l} \pi \,\Sigma \\ \bar{K}\, N \\ \eta\, \Lambda \\ K\, \Xi \\
\eta' \Lambda \\ \eta_c \Lambda \\ \bar{D}_s \Lambda_c \\ \bar{D}\, \Xi_c \\
\bar{D} \,\Xi_c'\end{array}$
& $\begin{array}{c} 1413 \\ 10 \end{array}$ & $\begin{array}{c} 0.7 \\ 2.7 \\ 1.1 \\ 0.1 \\ 0.0 \\ 0.0 \\ 0.2 \\ 0.0 \\ 0.0 \end{array}$
& $\begin{array}{c} 1413 \\ 10 \end{array}$ & $\begin{array}{c} 0.7 \\ 2.7 \\ 1.1 \\ 0.1 \\ 0.0 \\ 0.0 \\ 0.2 \\ 0.0 \\ 0.0 \end{array}$\\
\cline{3-7}
&$(0,-1)$    &
$\begin{array}{l} \pi\, \Sigma \\ \bar{K}\, N \\ \eta \,\Lambda \\ K\, \Xi \\ \eta' \Lambda \\
\eta_c \Lambda \\ \bar{D}_s \Lambda_c \\ \bar{D}\, \Xi_c \\ \bar{D}\, \Xi_c'\end{array}$
& $\begin{array}{c} 1689 \\ 35 \end{array}$ & $\begin{array}{c} 0.2 \\ 0.6 \\ 1.1 \\ 3.6 \\ 0.0 \\ 0.0 \\ 0.1 \\ 0.1 \\ 0.1 \end{array}$
& $\begin{array}{c} 1689 \\ 35 \end{array}$ & $\begin{array}{c} 0.2 \\ 0.6 \\ 1.1 \\ 3.6 \\ 0.0 \\ 0.0 \\ 0.1 \\ 0.1 \\ 0.1 \end{array}$\\
\cline{3-7}
&$(0,-1)$    &
$\begin{array}{l} \pi \,\Sigma \\ \bar{K}\, N \\ \eta\, \Lambda \\ K \,\Xi \\ \eta' \Lambda \\
\eta_c \Lambda \\ \bar{D}_s \Lambda_c \\ \bar{D}\, \Xi_c \\ \bar{D}\, \Xi_c'\end{array}$
& $\begin{array}{c} 3148 \\ 1.0  \end{array}$ & $\begin{array}{c} 0.04 \\ 0.03 \\ 0.03 \\ 0.04 \\ 0.08 \\ 0.08 \\ 3.2 \\ 5.0 \\ 0.1  \end{array}$
& $\begin{array}{c} 3234 \\ 0.57 \end{array}$ & $\begin{array}{c} 0.04 \\ 0.03 \\ 0.03 \\ 0.04 \\ 0.01 \\ 0.06 \\ 3.0 \\ 5.0 \\ 0.01 \end{array}$\\
\hline
\end{tabular}}
\label{tab:charm0a}
\end{center}
\end{table}

\begin{table}[t]
\begin{center}
\tbl{Continuation of Tab. \ref{tab:charm0a}.}
{\rescale \setlength{\tabcolsep}{1.0mm}
\setlength{\arraycolsep}{2.0mm}
\renewcommand{\arraystretch}{0.75}
\begin{tabular}{|ll|c|c|c|c|c|}
\hline $C=0:$ &$ (\,I,\phantom{+}S)$  &
$\rm state$ &  $\begin{array}{c} M_R [\rm MeV]  \\ \Gamma_R \,[\rm MeV]  \end{array}$ &
$|g_R|$  & $\begin{array}{c} M_R [\rm MeV]  \\ \Gamma_R \,[\rm MeV]  \end{array}$ & $|g_R|$  \\
\hline
\hline
&$(0,-1)$    &
$\begin{array}{l} \pi\, \Sigma \\ \bar{K}\, N \\ \eta \,\Lambda \\ K \,\Xi \\
\eta' \Lambda \\ \eta_c \Lambda \\ \bar{D}_s \Lambda_c \\ \bar{D} \,\Xi_c \\ \bar{D}\, \Xi_c'\end{array}$
& $\begin{array}{c} 3432 \\ 161 \end{array}$ & $\begin{array}{c} 0.1  \\ 0.0  \\ 0.0  \\ 0.1  \\ 1.3  \\ 0.7  \\ 0.6  \\ 0.1  \\ 5.6 \end{array}$
& $\begin{array}{c} 3581 \\ 4.9 \end{array}$ & $\begin{array}{c} 0.06 \\ 0.01 \\ 0.03 \\ 0.07 \\ 0.20 \\ 0.93 \\ 0.05 \\ 0.02 \\ 5.3 \end{array}$\\
\hline
&$(1,-1)$    &
$\begin{array}{l} \pi \,\Lambda \\ \pi\, \Sigma \\ \bar{K}\, N \\ \eta\, \Sigma \\ K \,\Xi \\
\eta' \Sigma \\ \eta_c \Sigma \\ \bar{D}\, \Xi_c \\ \bar{D}_s \Sigma_c \\ \bar{D}\, \Xi_c' \end{array} $
& $\begin{array}{c} 3602 \\ 227 \end{array}$ & $\begin{array}{c} 0.1   \\ 0.1  \\ 0.2  \\ 0.1  \\ 0.1  \\ 1.5  \\ 1.2 \\ 0.6  \\ 4.6 \\ 2.9 \end{array}$
& $\begin{array}{c} 3930 \\ 11  \end{array}$ & $\begin{array}{c} 0.08  \\ 0.04 \\ 0.12 \\ 0.08 \\ 0.06 \\ 0.27 \\ 1.8 \\ 0.11 \\ 3.6 \\ 2.4 \end{array}$ \\
\hline
&$(\frac12,-2)$  &
$\begin{array}{l} \pi\, \Xi \\ \bar{K}\, \Lambda \\ \bar{K}\, \Sigma \\ \eta\, \Xi \\ \eta' \Xi \\
\eta_c \Xi \\ \bar{D}_s \Xi_c \\ \bar{D}_s \Xi_c' \\ \bar{D}\, \Omega_c \end{array}$
& $\begin{array}{c} 1644 \\ 3.0 \end{array}$ & $\begin{array}{c}  0.1 \\ 0.4 \\ 2.8 \\ 1.3 \\ 0.0 \\ 0.0 \\ 0.2 \\ 0.1 \\ 0.0 \end{array}$
& $\begin{array}{c} 1644 \\ 3.1 \end{array}$ & $\begin{array}{c}  0.1 \\ 0.4 \\ 2.8 \\ 1.3 \\ 0.0 \\ 0.0 \\ 0.2 \\ 0.1 \\ 0.0 \end{array}$ \\
\cline{3-7}
&$(\frac12,-2)$  &
$\begin{array}{l} \pi\, \Xi \\ \bar{K}\, \Lambda \\ \bar{K}\, \Sigma \\ \eta\, \Xi \\ \eta' \Xi \\
\eta_c \Xi \\ \bar{D}_s \Xi_c \\ \bar{D}_s \Xi_c' \\ \bar{D}\, \Omega_c \end{array}$
& $\begin{array}{c} 3624 \\ 204 \end{array}$ & $\begin{array}{c} 0.1  \\ 0.1  \\ 0.1  \\ 0.0  \\ 1.4  \\ 1.0 \\ 0.6  \\ 3.3 \\ 4.3 \end{array}$
& $\begin{array}{c} 3798 \\ 6.0 \end{array}$ & $\begin{array}{c} 0.08 \\ 0.04 \\ 0.04 \\ 0.01 \\ 0.22 \\ 1.2 \\ 0.10 \\ 2.9 \\ 4.0 \end{array}$ \\
\hline
\end{tabular}}
\label{tab:charm0b}
\end{center}
\end{table}

The states are narrow as a result of the OZI rule. The mechanism
is analogous to the one explaining the long life time of the
$J/\Psi$-meson. We should mention, however, a caveat. It turns out
that the width of the crypto-exotic states is quite sensitive to
the presence of channels involving the $\eta'$ meson. This is a
natural result since the $\eta'$ meson is closely related to the
$U_A(1)$ anomaly giving it large gluonic components. The latter
work against the OZI rule. We emphasize that switching off the
t-channel exchange of charm or using the SU(4) estimate for the
latter, strongly bound crypto-exotic states are formed. In Tabs.
\ref{tab:charm0a}-\ref{tab:charm0b} the zero-charm spectrum
insisting on the SU(4) estimates (\ref{meson-SU4-result},
\ref{baryon-SU4-result}) is shown in the 3rd and 4th column. The
mass of the crypto-exotic nucleon resonance comes at 3.33 GeV in
this case. Its width of 160 MeV is completely dominated by the
$\eta' N$ decay. The properties of that state can be adjusted
easily to be consistent with the empirical values claimed in
\cite{Karnaukhov:91}. The $\eta'$ coupling strength to the
open-charm mesons can be turned off by decreasing the magnitude of
$h^{\bar 3}_{ \bar 3 1 }$ and $h^{\bar 3}_{ 1\bar 3 }$ by 33.3
$\%$ away from their SU(4) values. As a result the width of the
resonance is down to about 1-2 MeV. It is stressed that the masses
of the crypto-exotic states are not affected at all. The latter
are increased most efficiently by allowing an OZI violating
$\phi_\mu \,D \bar D $ vertex. We adjust $h_{\bar 3 \bar 3}^1
\simeq -1.19 \,g$  and $h^{\bar 3}_{ \bar 3 1 }=h^{\bar 3}_{1 \bar
3  } \simeq 0.71 \,g $ as to obtain the nucleon resonance mass and
width at $3.52$ GeV and $7$ MeV. For all other parameters the
SU(4) estimates are used. The result of this choice of parameters
is shown in the last two rows of Tabs.
\ref{tab:charm0a}-\ref{tab:charm0b}. Further  crypto-exotic
states, members of the aforementioned octet, are predicted at mass
3.58 GeV $(0,-1)$ and 3.93 GeV $(1,-1)$. The multiplet is
completed with a $(\frac{1}{2},-2)$ state at 3.80 GeV. The decay
widths of these states center around $\sim 7$ MeV. This  reflects
the dominance of their decays into channels involving the $\eta'$
meson. The coupling constants to the various channels are included
in Tabs. \ref{tab:charm0a}-\ref{tab:charm0b}. They confirm the
interpretation that the crypto-exotic states discussed above are a
consequence of a strongly attractive force between the charmed
mesons and the baryon sextet.

We close with a discussion of the crypto-exotic SU(3) singlet
state, which is formed due to strong attraction in the $(\bar
D_s\Lambda_c), (\bar D\,\Xi_c )$ system. Its nature is quite
different as compared to the one of the octet states. This is
because its coupling to the $\eta' \Lambda$ channel is largely
suppressed. Indeed its width is independent on the magnitude of
$h^{\bar 3}_{ \bar 3 1 }=h^{\bar 3}_{1 \bar 3  }$ as demonstrated
in Tabs. \ref{tab:charm0a}-\ref{tab:charm0b}. We identify this
state with a signal claimed in the $K^-p$ reaction, where a narrow
hyperon state with 3.17 GeV mass and width smaller than 20 MeV was
seen \cite{Amirzadeh:79}. Using values for the coupling constants
as suggested by SU(4) the state has a mass and width of 3.148 GeV
and 1 MeV (see 3rd and 4th column of Tabs.
\ref{tab:charm0a}-\ref{tab:charm0b}). Our favored parameter set
with $h_{\bar 3 \bar 3}^1  \simeq -1.19 \,g$  and $h^{\bar 3}_{
\bar 3 1 }=h^{\bar 3}_{1 \bar 3  } \simeq 0.71 \,g $ predicts a
somewhat reduced binding energy.

\section{Summary}

We reviewed a coupled-channel study of s-wave baryon resonances
with charm $0$. The interaction is defined by the exchange of
light vector mesons in the t-channel. All relevant coupling
constants are obtained from chiral and large-$N_c$ properties of
QCD. Less relevant vertices related to the t-channel forces
induced by the exchange of charmed vector mesons  were estimated
by applying SU(4) symmetry. Most spectacular is the prediction of
narrow crypto-exotic baryons with charm zero forming below 4 GeV.
Such states contain a $c \bar c$ pair. Their widths parameters are
small due to the OZI rule, like it is the case for the $J/\Psi$
meson. We predict an octet of crypto-exotic states which decay
dominantly into channels involving an $\eta'$ meson. An even
stronger bound crypto-exotic SU(3) singlet state is predicted to
have a decay width of about 1 MeV only. We recover the masses and
widths of a crypto-exotic nucleon and hyperon resonance suggested
in high statistic bubble chamber experiments
\cite{Amirzadeh:79,Karnaukhov:91}.

\end{document}